\documentclass[preprint,showpacs,aps,preprintnumbers,amsmath,amssymb,
nofootinbib]{revtex4}

\usepackage{epsfig}
\usepackage{graphicx}
\begin{document}

\begin{flushright}
\end{flushright}


\newcommand{\be}{\begin{equation}}
\newcommand{\ee}{\end{equation}}
\newcommand{\bea}{\begin{eqnarray}}
\newcommand{\eea}{\end{eqnarray}}
\newcommand{\bers}{\begin{eqnarray*}}
\newcommand{\eers}{\end{eqnarray*}}
\newcommand{\nn}{\nonumber}
\newcommand\un{\cal{U}}
\def\Apa{A_\parallel}
\def\Ape{A_{\perp}}
\def\lp{\lambda^\prime}
\def\ll{\Lambda}
\def\mb{m_{\Lambda_b}}
\def\ml{m_\Lambda}
\def\s1{\hat s}
\def\ds{\displaystyle}
\def\s{\smallskip}
\def\l{\hspace*{0.05cm}}
\def\esp{\hspace*{1cm}}

\title{\large Explaining $B \to K \pi $ anomaly 
with non-universal $Z'$ boson}
\author{R. Mohanta$^1$  and A. K. Giri$^2$  }
\affiliation{$^1$ School of Physics, University of Hyderabad,
Hyderabad - 500 046, India\\
$^2$ Department of Physics, Punjabi University,
Patiala - 147 002, India}

\begin{abstract}
We study the effect of non-universal $Z'$ boson in the decay modes
$B \to K \pi$. In the standard model these modes receive
dominant contributions from $b \to s$ QCD penguins. Therefore, in
this limit one expects $S_{\pi^0 K^0} \approx \sin 2 \beta $,
$A_{ \pi^0 K^0} \approx 0$ and $A_{ \pi^0 K^-} \approx
A_{\pi^+ K^-}$. The corrections due to the presence of
small non-penguin contributions is found to yield $S_{ \pi^0 K^0}
> \sin 2 \beta $ and $\Delta A_{CP}(K \pi) \simeq 2.5 \%$.
 However, the measured value of $S_{ \pi^0 K^0}$ is
less than $ \sin 2 \beta $ and $\Delta A_{CP}(K \pi) \simeq 15 \%$. 
We show the model with a non-universal
$Z'$ boson can successfully explain these anomalies.

\end{abstract}

\maketitle

The standard model (SM) of electroweak interaction is very
successful in explaining the observed data so far, but still it is
believed that there must exist some new physics beyond the SM, whose
true nature is not yet well-known. Therefore, intensive search for
physics beyond the SM is now being performed in various areas of
particle physics. In this context the $B$ system can also be used as
a complementary probe. One of the important ways to
look for new physics in the $b$-sector is the analysis of rare $B$
decay modes, which are induced by the flavor changing neutral
currents (FCNCs), in particular $b \to s$ transitions. Although, so
far we have not been able to see any clear indication of physics
beyond the SM in the currently running $B$-factories but there
appears to be some kind of deviation in some $b\to s$ penguin
induced transitions i.e., the mixing induced CP asymmetries in many
$b \to s \bar q q$ penguin dominated modes do not seem to agree with
the SM expectations. The measured values in such modes 
follow the trend $S_{s
\bar q q} < \sin 2 \beta $ \cite{hfag}, whereas in the SM they  are
expected to be similar \cite{yg}. In this context $B \to K \pi$
decay modes, which receive dominant contributions from $b \to s$
mediated QCD penguins in the SM, provide an ideal testing
ground to look for new physics.

At present, there seems to be two possible hints of new physics
in these modes. The first one is associated with the mixing induced
CP asymmetry in $B^0 \to \pi^0 K^0$ mode. 
 The time dependent CP asymmetry in this mode is defined
as \bea \frac{\Gamma(\bar B^0(t) \to \pi^0 K_s)-\Gamma(B^0(t) \to
\pi^0 K_s)}{\Gamma(\bar B^0(t) \to \pi^0 K_s)+\Gamma(B^0(t) \to
\pi^0K_s)}= A_{\pi^0 K_s} \cos(\Delta M_d t)+S_{\pi^0 K_s}
\sin(\Delta M_d t)\;, \eea and in the pure QCD penguin limit one
expects $A_{\pi^0 K_s} \approx 0$ and $S_{\pi^0 K_s} \approx \sin(2
\beta)$. Small non-penguin contributions do provide some corrections
to these asymmetry parameters and it has been shown in Refs.
\cite{mg, beneke, soni, hiller} that these corrections
generally tend to increase $S_{K
\pi^0}$ from its pure penguin limit of ($\sin 2 \beta$) by a
modest  amount i.e., $ S_{\pi^0 K_s} \approx 0.8$.
Recently, using isospin symmetry it has been shown in \cite{rf, mg1}
that the standard model favors a large $S_{\pi^0 K_s} \approx 0.99$.

However, the  recent results from Belle
\cite{belle} and Babar \cite{babar}  are \bea A_{\pi^0 K_s}&=& 
0.14 \pm 0.13 \pm
0.06,~~~~~S_{\pi^0 K_s}=0.67 \pm 0.31 \pm 0.08~~~~({\rm
Belle})\nn\\
A_{\pi^0 K_s}&=& -0.13 \pm 0.13 \pm 0.03,~~~~~S_{\pi^0 K_s}=0.55 \pm
0.20 \pm 0.03~~~~({\rm Babar}) \eea with average \be A_{\pi^0 K_s}=
-0.01 \pm 0.10,~~~~~S_{\pi^0 K_s}=0.57 \pm 0.17\;, \ee 
where $S_{\pi^0 K_s}$ is found to be
smaller than the present  world average value of $\sin 2 \beta=0.672
\pm 0.024 $ measured in $b \to c \bar c s$ transitions 
\cite{hfag} by
nearly 1-sigma.

This deviation which is  opposite to the SM expectation, implies the
presence of new physics in the $ B^0 \to K^0 \pi^0 $ decay
amplitude. In the SM, this decay mode receives contributions from
QCD penguin ($P$), electroweak penguin ($P_{EW})$ and color
suppressed tree ($C$) diagrams, which follow the hierarchical
pattern $P:P_{EW}:C= 1:\lambda : \lambda^2$, where $\lambda \approx
0.2257$ is the Wolfenstein expansion parameter. Thus, accepting the
above discrepancy seriously one can see that the electroweak penguin
sector is the best place to search for new physics.

 The second anomaly in the $B \to K \pi$ sector is associated
with the  direct
CP asymmetry parameters of $B^-\to \pi^0K^- $ and that of the $ \bar
B^0\to \pi^+K^-$. The $\Delta A_{CP}(K \pi)$ puzzle refers to the 
difference in direct
CP asymmetries in $B^- \to
\pi^0 K^-$ and $\bar B^0 \to \pi^+ K^-$ modes. These two modes
receive similar dominating contributions from tree and QCD penguin
diagrams and hence one would naively expect that these two channels
will have similar direct CP asymmetries i.e., ${ A}_{\pi^0
K^-} \approx { A}_{\pi^+ K^-}$. In the QCD factorization approach, the
difference between these asymmetries  is found to be \cite{soni10}
 \be \Delta A_{CP} = A_{ K^- \pi^0} - A_
{ K^- \pi^+} =(2.5 \pm
 1.5)\%
 \ee
 whereas the corresponding experimental value is \cite{hfag}
\be \Delta A_{CP} =(14.8 \pm
 2.8)\%\;,
 \ee
which yields nearly  $4 \sigma$ deviation.
This constitutes what is called 
$\Delta A_{CP}(K \pi )$ puzzle in the literature and 
is believed to be an indication of
the existence of new physics.

To account for these discrepancies between the observed and expected
observables, here we consider the effect due to an extra $U(1)'$
gauge boson $Z'$ as an illustration, which can provide additional
contributions to the electroweak penguin sector. The existence of
extra $Z'$ boson is a  feature of many models addressing physics
beyond the SM, e.g., models based on extended gauge groups
characterized by additional $U(1)$ factors \cite{zp1}. Also the new
physics models which contain exotic fermions, predict the existence
of additional gauge boson \cite{e6}. Flavor mixing can be induced at
the tree level in the up-type and/or down-type quark sector after
diagonalizing their mass matrices.

Before incorporating the effect of extra $Z'$ boson to the $B \to
K \pi$ amplitudes, first we would like to briefly present  the
standard model results. In the SM, the relevant effective
Hamiltonian describing the decay modes $\bar B \to \pi 
\bar K $ is given
by
\begin{equation}
{\cal H}_{eff}^{SM} = \frac{G_F}{\sqrt{2}}\left[  V_{ub}
V_{us}^*(C_1O_1+C_2 O_2)- V_{tb}V_{ts}^*\sum_{i=3}^{10} C_i O_i
\right].
\end{equation}

Thus, one can obtain the transition amplitudes in the QCD
factorization approach \cite{qcdf} as  \bea 
\sqrt 2 A(\bar B^0 &\to & \pi^0
\bar K^0)=\lambda_u A_{\bar
K \pi}\alpha_2 \nn\\
&+&\sum_{p=u,c}\lambda_p\Big[A_{\pi \bar
K}\Big(-\alpha_4^p+\frac{1}{2}\alpha_{4,EW}^p-
\beta_3^p+\frac{1}{2}\beta_{3,EW}^p \Big)
+\frac{3}{2}A_{\bar K \pi}\alpha_{3,EW}^p \Big]\;, 
\label{smamp}\eea 
 \bea \sqrt 2 A(B^- \to
\pi^0 K^-)&=&\lambda_u \Big(A_{\pi \bar K}(\alpha_1+\beta_2)+A_{\bar
K \pi}\alpha_2 \Big)\nn\\
&+&\sum_{p=u,~c}\lambda_q\Big(A_{\pi \bar
K}(\alpha_4^p+\alpha_{4,EW}^p+\beta_3^p+\beta_{3,EW}^p)+\frac{3}{2}A_{\bar
K \pi}\alpha_{3,EW}^p \Big) \eea and \bea A(\bar B^0 \to  \pi^+
K^-)=\lambda_u ~A_{\pi \bar K}~\alpha_1 +
\sum_{p=u,~c}\lambda_q A_{\pi
\bar
K}\Big(\alpha_4^p+\alpha_{4,EW}^p+\beta_3^p-\frac{1}{2}
\beta_{3,EW}^p \Big),\label{sma1}
\eea
 where $\lambda_p= V_{pb}V_{ps}^*$ and
\be  A_{\pi \bar K}= i\frac{G_F}{\sqrt 2}M_B^2 F_0^{B\to
\pi}f_K,~~~~~~~~A_{ \bar K \pi}= i\frac{G_F}{\sqrt 2}M_B^2
F_0^{B\to K}f_{\pi}\;. \ee
The parameters $\alpha_i$'s and $\beta_i$'s are related to the
Wilson coefficients $C_i$'s and the corresponding expressions can be
found in \cite{qcdf}.

Thus, one can symbolically represent this amplitude as \bea
 A(\bar B \to \pi \bar K)= \lambda_u A_u +\lambda_c A_c
=\lambda_c A_c\Big[1+ r~ a~ e^{i(\delta_1-\gamma)} \Big],
\label{amp}\eea
 where $a=|\lambda_u/\lambda_c|$,
$-\gamma$ is the weak phase of $V_{ub}$,  $r=|A_u/A_c|$, and
$\delta_1$  is the relative strong phases between $A_u$ and $A_c$.
From the above amplitude, the CP averaged branching
ratio, direct and mixing induced  CP asymmetry
(for neutral $B$ meson case) parameters
can be obtained as \bea
{\rm Br} &=& \frac{|p_{c.m}| \tau_B}{8 \pi M_B^2}
|\lambda_c A_c|^2(1+(r a)^2+2 r a \cos \delta_1 \cos \gamma)\;,\nn\\
 A_{\pi K}&=&\frac{2r~a \sin \delta_1
\sin \gamma } {1+(r~a)^2+2ra \cos \delta_1 \cos
\gamma }\;,\nn\\
S_{\pi K}&= &\frac{ \sin 2 \beta+ 2 r a \cos \delta_1 \sin(2
\beta+\gamma) +(r a)^2\sin(2 \beta +2 \gamma) } {1+(ra)^2+2ra \cos
\delta_1 \cos \gamma}\;.
 \eea

For numerical evaluation,  we use input parameters as given in the
S4 scenario of QCD factorization approach \cite{qcdf}. For the CKM matrix
elements we use $V_{cb}=(41.5_{-1.1}^{+1.0})
 \times 10^{-3}$, $|V_{cs}|=0.97334\pm 0.00023$,
$|V_{ub}|=(3.59\pm0.16) \times 10^{-3}$, $|V_{us}|=0.2257\pm 0.0010
$ \cite{pdg} and $\gamma=\left (65 \pm 10 \right )^\circ$. The
particle masses and life time of $B^0$ are taken from \cite{pdg}.
Since QCD factorization suffers from end-point divergences
we have included $20\%$ uncertainties in the branching ratio and
10 $\% $ uncertainties in the CP asymmetry parameters.
With these inputs we show in Figure-1 the correlation plots
between the branching ratio and $S_{\pi^0 K_s}$ (left panel) and
between $S_{\pi^0 K_s}$ and $A_{\pi^0 K_s}$ (right panel). From
these figures it can be seen that the obtained value $A_{\pi^0 K_s}$
is in accordance with the SM expectation. However,
although the SM result of
$S_{\pi^0 K_s}$ lies within its observed $1-\sigma $ range, but the
branching ratio is well below the
corresponding observed value. Hence the present situation is that,
it appears difficult to accommodate simultaneously these three observables
in the standard model.

\begin{figure}[htb]
\includegraphics[width=8cm,height=6cm, clip]{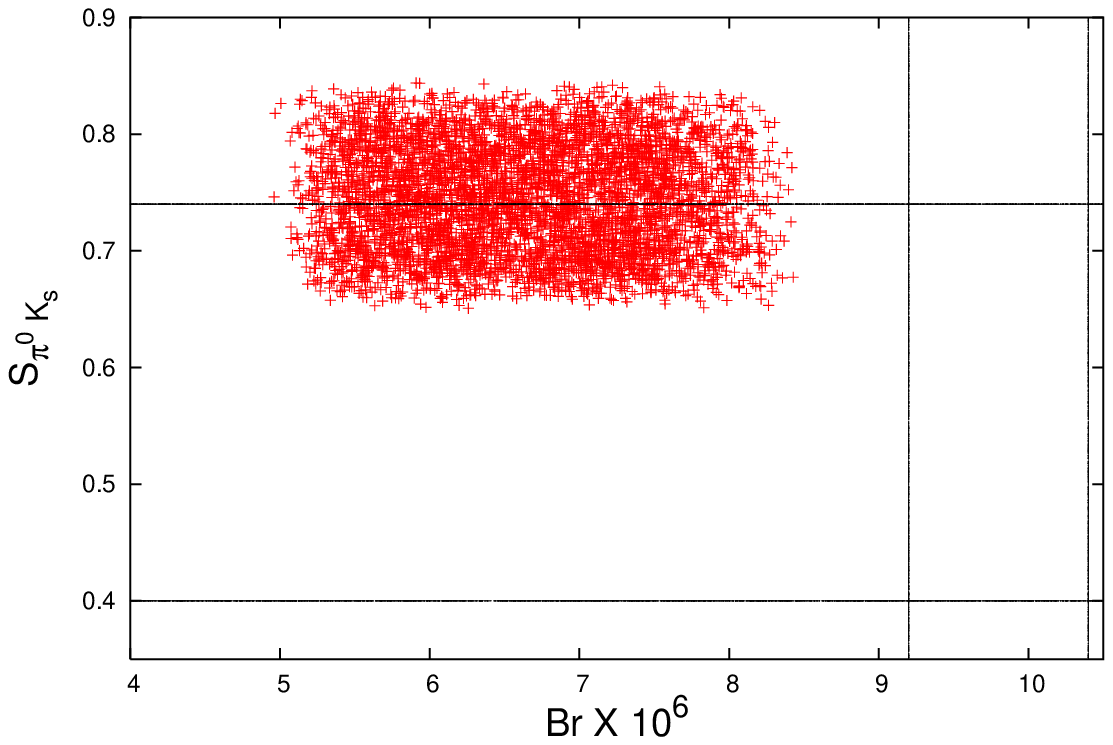}
\hspace{0.2 cm}
\includegraphics[width=8cm,height=6cm, clip]{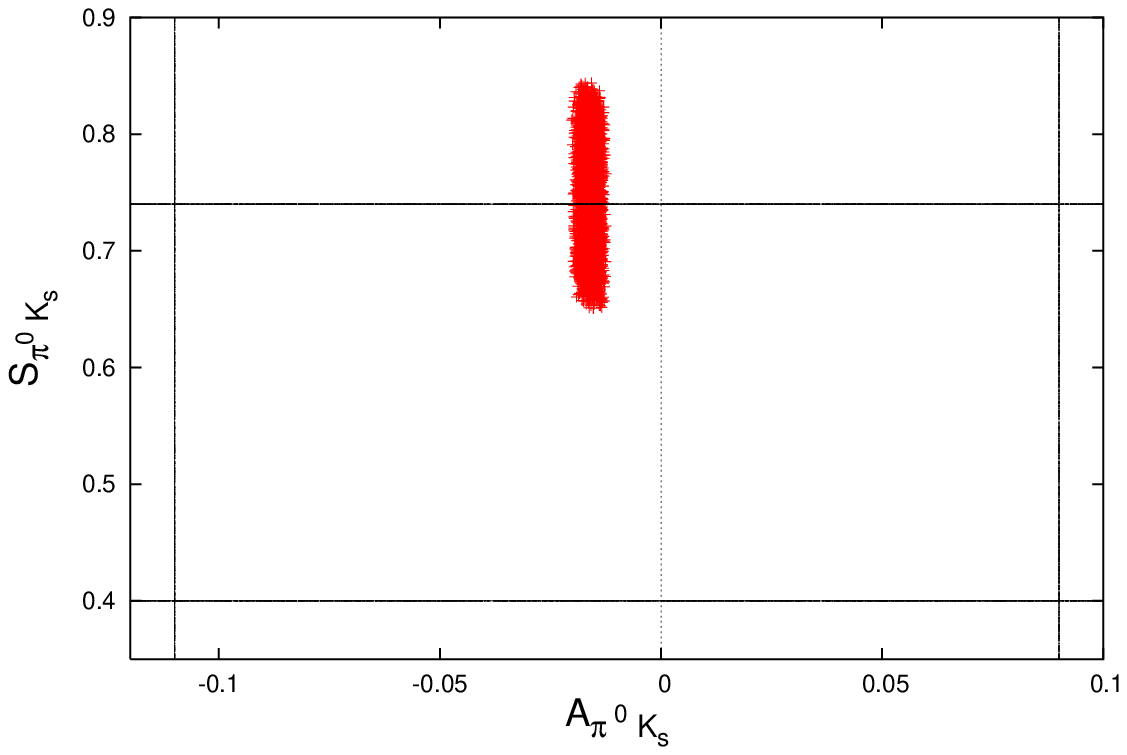}
\caption{Correlation plots  between the mixing induced CP asymmetry
$S_{\pi^0 K_s}$ and the CP averaged BR (left panel) and $S_{\pi^0
K_s}$ and the direct CP asymmetry $A_{\pi^0 K_s}$  (right panel) for
the $B^0 \to \pi^0 K_s$ in the Standard Model. The horizontal and
vertical lines in both the figures represent the $1-\sigma$ allowed
rages of the respective observables. }
\end{figure}

Now we will consider the effect of the non-universal $Z'$ boson on
these decay modes $B \to \pi K$. As discussed earlier, the FCNCs
due to $Z'$ exchange can be induced by mixing among the SM quarks
and the exotic quark which have different $Z'$ quantum numbers.
 Here we will consider
the model in which the interaction between the $Z'$ boson and
fermions are flavor nonuniversal for left handed couplings and
flavor diagonal for right handed couplings. The detailed description
of the family nonuniversal $Z'$ model with flavor changing neutral
currents can be found in Ref. \cite{zp1}. 
 The search for the extra
$Z'$ boson occupies an important place in the experimental programs
of the Fermilab Tevatron and CERN LHC \cite{zp3}. The implications
of the FCNC mediated $Z'$ boson effect has been extensively studied
in the context of $B$ physics \cite{zp2,zp2a,zp2b,desh,rm}.

The effective Hamiltonian describing the transition $b \to 
s \bar q q$, where $q=u,d$ for $\bar B \to \pi \bar K$, mediated by the
$Z'$ boson is given by \cite{zp2a} \be {\cal H}_{eff}^{Z'}= \frac{2
G_F}{\sqrt 2} \left ( \frac{g' M_Z} {g_1 M_{Z'}} \right )^2
B_{sb}^{L}~(\bar s b)_{V-A}\sum_q \left [(B_{qq}^L~ (\bar q q)_{V-A}
+B_{qq}^R~ (\bar q q)_{V+A}\right ], \label{zp0} \ee
 where $g_1=e/(\sin \theta_W \cos \theta_W)$, $g'$ is the gauge
 coupling constant of the $U(1)'$ group  and
$B_{ij}^{L(R)}$ denote the left (right) handed effective $Z'$
couplings of the quarks $i$ and $j$ at the weak scale. The diagonal
elements are real due to the hermiticity of the effective
Hamiltonian but the off diagonal elements may contain effective weak
phase. Therefore, both the terms in (\ref{zp0}) will have the same
weak phase due to $B_{sb}^L$.

Since the  structure of the effective Hamiltonian (\ref{zp0}) in
this model has the same form as that of the SM, the effect of $Z'$
can be represented as a modification of the SM Wilson coefficients
of the corresponding operators. Assuming that $B_{uu}^{L(R)} \simeq
-2B_{dd}^{L(R)}$, so that the new physics is primarily manifest in
the EW penguins \cite{zp2a}, the resulting effective Hamiltonian at
the $M_W$ scale is given as \bea {\cal H}_{eff}^{Z'}&= &- \frac{4
G_F}{\sqrt 2}\left ( \frac{g' M_Z} {g_1 M_{Z'}} \right )^2
B_{sb}^{L}\sum_q \left
[(B_{dd}^L~ O_9^{(q)} +B_{dd}^R~ O_7^{(q)}\right ]\nn\\
&= &- \frac{ G_F}{\sqrt 2}V_{tb}V_{ts}^*\sum_q \left [\Delta C_9
O_9^{(q)} + \Delta C_7 O_7^{(q)}\right ], 
\label{zp} \eea where $\Delta C_7$ and $\Delta C_9$ are
the new Wilson coefficients arising due to the extra $Z'$ boson
contributing to the electroweak penguin sector given as
 \be \Delta C_9= \left ( \frac{g'
M_Z} {g_1 M_{Z'}} \right )^2 \left ( \frac{B_{sb}^{L}
B_{dd}^L}{V_{tb}V_{ts}^*}\right )\;,~~~ \Delta C_7= \left ( \frac{g'
M_Z} {g_1 M_{Z'}} \right )^2 \left (\frac{B_{sb}^{L}
B_{dd}^R}{V_{tb} V_{ts}^*}\right ). \ee For convenience we can
parameterize these coefficients as
 \be \Delta C_9\equiv\xi_L= -|\xi_L| e^{i \phi_s}\;,~~~
\Delta C_7\equiv \xi_R=-|\xi_R| e^{i \phi_s}\;, \ee
where $\phi_s$ is the weak phase associated with $B_{sb}^L$
and the minus signs in the r.h.s appear because the weak phase
of the CKM element $V_{ts}$ is $\pi$.

After having an idea about the new electroweak penguin contributions
at the $M_Z$ scale, we  now evolve them to the $b$ scale
using renormalization group equation \cite{wilson}.  Using the
values of these coefficients at $b$ scale we obtain the new
contribution to the transition amplitudes. These values  can then be
evolved to the $m_b$ scale using the renormalization group equation
\cite{wilson}
\begin{equation}
{\vec{C}} (m_b) = U_{5} (m_{b},M_{W}, \alpha)~ {\vec C} (M_{W}),
\end{equation}
where $C$ is the $10 \times 1$ column vector of the Wilson
coefficients and $U_{5}$ is the five flavor $10 \times 10$ evolution
matrix. The explicit forms of ${\vec C}(M_W)$ and $U_{5} (m_b, M_W,
\alpha)$ are given in \cite{wilson}.

Because of the RG evolution these three Wilson coefficients generate
new set of Wilson coefficients $\Delta C_i (i=3,\cdots, 10)$ at the
low energy regime (i.e., at the $m_b$ scale) as presented in
Table-1.

\begin{table}[htbp]
\begin{center}
\begin{tabular}{|c|c|c|c|}
\hline $\Delta C_3$ & $\Delta C_4$ & $\Delta C_5$&$\Delta C_6$\\
\hline  $0.05 ~\xi_L - 0.01~\xi_R $ &$-0.14 ~\xi_L + 0.008~\xi_R
$&$0.029~
\xi_L - 0.017~\xi_R $&$-0.162~ \xi_L + 0.01~\xi_R $\\
\hline \hline
$\Delta C_7$&$\Delta C_8$&$\Delta C_9$&$\Delta C_{10}$ \\
\hline $0.036 ~\xi_L - 3.65~\xi_R $&$0.01~ \xi_L - 1.33~\xi_R
$&$-4.41 ~\xi_L + 0.04~\xi_R
$&$0.99~ \xi_L - 0.005~\xi_R $\\
\hline
\end{tabular}
\caption{Values of the new Wilson coefficients at the $m_b$ scale.
 } \label{tab2}
\end{center}
\end{table}

Thus one can obtain the new contribution to the transition amplitude
\bea \sqrt 2 A(\bar B^0 \to  \pi^0 \bar K^0) &= &A^{SM}_{\pi^0 K^0}
 -\lambda_t \Big(A_{\pi
\bar K}(-\Delta
\alpha_4+\frac{1}{2}\Delta\alpha_{4,EW}-\Delta\beta_3+\frac{1}{2}
\Delta\beta_{3,EW})\nn\\
& +&\frac{3}{2}A_{\bar K \pi}\Delta\alpha_{3,EW}
\Big)\;, \eea
 \bea \sqrt2 A(B^- \to
\pi^0 K^-)&=& A^{SM}_{\pi^0 K^-}-\lambda_t 
\Big (\frac{3}{2}A_{\bar
K \pi}\Delta\alpha_{3,EW}\nn\\
&-&A_{\pi \bar K}(\Delta \alpha_4+\Delta
\alpha_{4,EW}+\Delta \beta_3+\Delta \beta_{3,EW}) \Big) \eea 
and \bea A(\bar B^0 \to  \pi^+
K^-)&=&A^{SM}_{\pi^- K^+}- \lambda_t~ A_{\pi \bar K}\Big(\Delta 
\alpha_4+\Delta
\alpha_{4,EW}+\Delta \beta_3-\frac{1}{2}\Delta \beta_{3,EW} \Big),
\eea 
 where $A^{SM}$'s are the corresponding SM amplitudes as
given in Eqs. (\ref{smamp}-\ref{sma1}) and  
$\Delta \alpha_i$'s and $\Delta \beta_i$'s are
related to the new Wilson coefficients $\Delta C_i$'s.
Analogous to Eq. (\ref{amp}) we can now represent the transition
amplitude incorporating the new physics contribution as
 \bea A(\bar B \to \pi \bar
K)= A^{SM}-\lambda_t ~\xi ~A^{N}= \lambda_c A_c\Big[1+
 r a~
e^{i(\delta_1-\gamma)}-r' b~ e^{i(\delta_2+\phi_s)}], \eea
 where for
simplicity we have assumed $\xi_L=\xi_R=\xi$. $A^N$ is the new
physics contribution which contains the strong phase information,
$b=|\lambda_{t}~ \xi/\lambda_c|$, $r'=|A^{N}/A_c|$, and $\delta_2$
is the relative strong phases between $A^N$ and $A_c$. Thus from the
above amplitude one can obtain the CP averaged branching ratio,
direct and mixing induced CP asymmetry parameters as
\bea
{\rm Br}&=& \frac{|p_{c.m}| \tau_B}{8 \pi M_B^2}
\Big[{\cal{R}}+2ra \cos \delta_1 \cos
\gamma-2r' b \cos \delta_2 \cos \phi_s - 2 r r' a b
 \cos(\delta_2-\delta_1)\cos(\gamma+\phi_s)\Big]\;,\nn\\
A_{\pi K}&=&\frac{2\Big[ra \sin
\delta_1 \sin \gamma +r' b \sin \delta_2 \sin \phi_s + r r' a b
\sin(\delta_2-\delta_1) \sin
(\gamma+\phi_s)\Big]}{\Big[{\cal{R}}+2ra \cos \delta_1 \cos
\gamma-2r' b \cos \delta_2\cos \phi_s  - 2 r r' a b
 \cos(\delta_2-\delta_1)\cos(\gamma+\phi_s)\Big]}\;,\nn\\
S_{\pi K}&= &\frac{X}{{\cal R}+2ra \cos \delta_1 \cos \gamma-2r'
b  \cos \delta_2 \cos \phi_s- 2 r r' a b\cos(\delta_2-\delta_1)
 \cos(\gamma+\phi_s)
}\;,
 \eea
where
${\cal{R}}=1+(ra)^2+(r'b)^2$ and \bea X &=& \sin 2 \beta+ 2 r a \cos
\delta_1 \sin(2 \beta+\gamma)- 2 r' b \cos \delta_2
\sin(2 \beta- \phi_s) +(r a)^2\sin(2 \beta +2 \gamma) \nn\\
&+&(r' b)^2\sin(2 \beta -2 \phi_s)-2 rr' a b \cos(\delta_2-
\delta_1) \sin(2 \beta+\gamma -\phi_s). \eea

 In order to see the effect of $Z'$ boson, we
have to know the values of the $\xi$  or equivalently $B_{sb}^L$ and
$B_{dd}^{L,R}$. Generally one expects $g'/g_1 \sim 1 $, if both the
$U(1)$ gauge groups have the same origin from some grand unified
theories. It has been shown in \cite{zp2a, zp2b} that the mass
difference of $B_s - \bar B_s $ mixing and the CP asymmetry anomaly
in  $B \to \phi K, \pi K $ can be resolved if $|B_{sb}^L
B_{ss}^{L,R}| \sim |V_{tb} V_{ts}^*|$. Assuming the universality of
first two generations one can obtain $B_{dd}^{L,R}\simeq
B_{ss}^{L,R}$ and hence $|\xi| \approx (M_Z/M_{Z'})^2$. The $Z'$
mass is constrained by direct searches at Fermilab, weak neutral
current data and precision studies at LEP \cite{mass}, which give a
model dependent lower bound around 500 GeV. Here we consider more
conservative limit as $M_{Z'} \geq 700 $ GeV, which gives the upper
bound of the parameters $|\xi| < 1.65 \times 10^{-2}$. However, in
this analysis we vary its value within the range $(0.015-0.005)$,
which is true for a TeV range $Z'$ boson. For a heavy massive $Z'$
boson (say $M_{Z'} > $1.5 TeV) the new physics contributions is
found to be very small and it does not have much impact on the SM
results.

Now varying $|\xi|$ in the range $0.005 \leq |\xi| \leq 0.015$, we
show in Figure-2 the correlation plots between $S_{\pi^0 K_s}$ and
the branching ratio (left panel) and $S_{\pi^0K_s}$ and $A_{\pi^0
K_s} $ in the right panel. The allowed region in $\Delta
A_{CP}(K \pi)$ and $|\xi|$ plane is shown in the figure-3.
 From these figures it can be seen that
the branching ratio and the CP violating parameters
in $B^0 \to \pi^0 K^0$ mode and the
 $\Delta A_{CP}(K\pi)$ puzzle can be
simultaneously explained in the model with an extra $Z'$ boson.

\begin{figure}[t]
\includegraphics[width=8cm,height=6cm,clip]{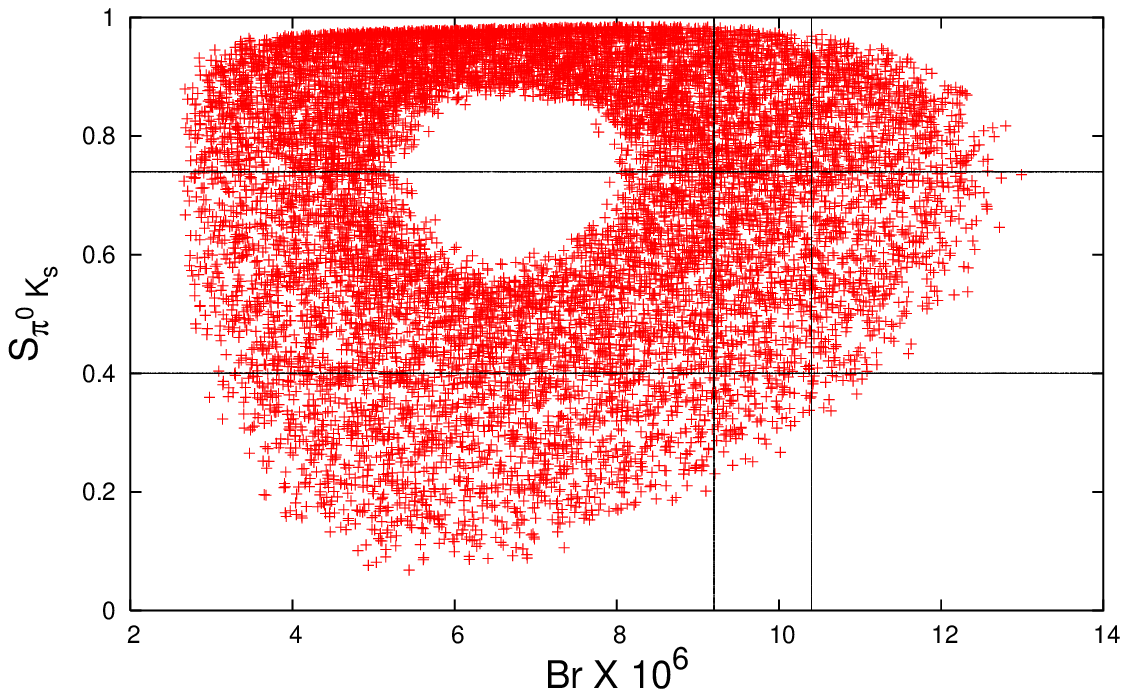}%
\hspace{0.2cm}%
\includegraphics[width=8cm,height=6cm,clip]{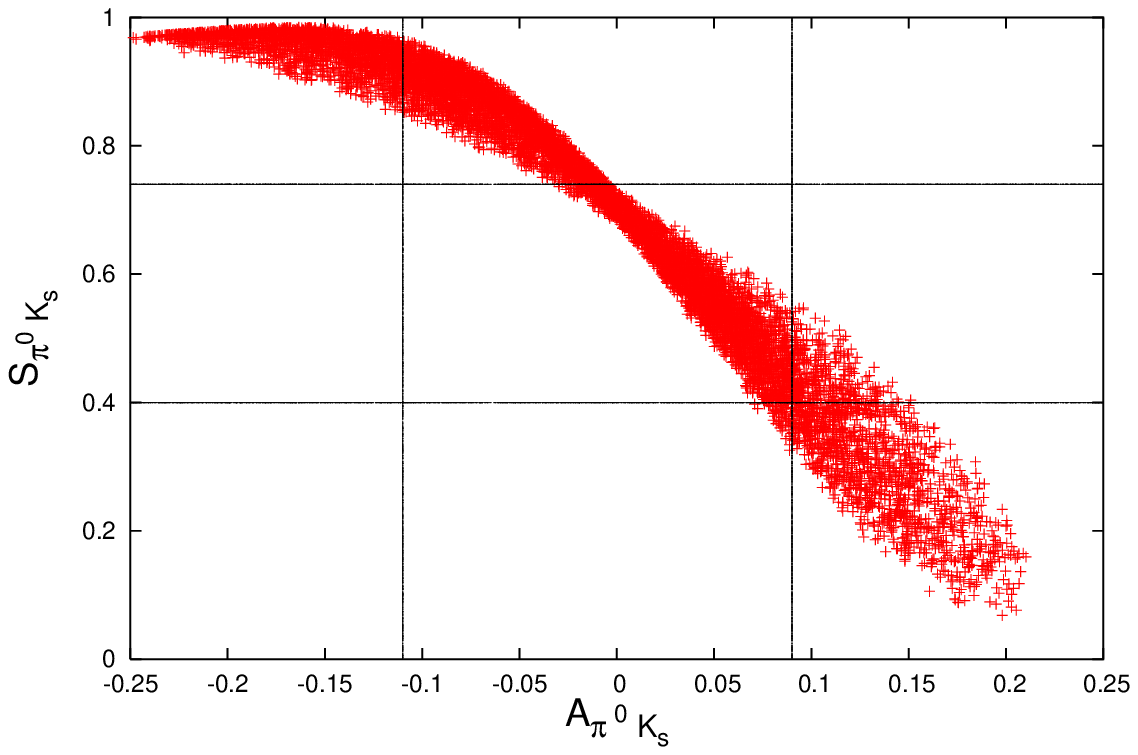}
\caption{Correlation Plots   (a) between the mixing induced CP
asymmetry $S_{\pi^0 K_s}$ and the direct CP asymmetry $A_{\pi^0
K_s}$ (b) between CP averaged BR and $S_{\pi^0 K_s}$ for the $B^0
\to \pi^0 K_s$ in the model with an extra $Z'$ boson. The horizontal
and vertical lines represent 1-sigma experimental allowed ranges.}
\end{figure}

\begin{figure}[htb]
\centerline{\epsfysize 2.5 truein \epsfbox{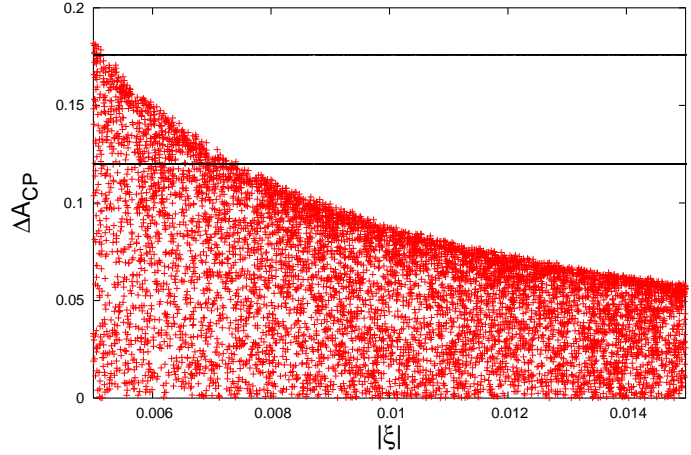}} \caption{The
allowed region  of the CP asymmetry difference ($\Delta A_{CP}$) in
the ($\Delta A_{CP}-|\xi|$) plane.
 The horizontal lines correspond to the experimentally
allowed $1-\sigma$ range. }
\end{figure}

It is well known that strangeness changing charmless $B$ decays,
dominated by $b \to s$ penguin amplitudes, which arise in the SM at
one-loop level, are highly sensitive to new physics effects.
Possible existence of new physics in these modes is being
intensively searched via the measurement of time dependent CP
asymmetries of neutral $B$ meson decays into final CP eigen states.
Virtual new heavy particles with mass scale typically around a TeV
range may affect the SM predictions $A_{CP} \approx 0$ and
$S_{CP}\approx \eta_{CP} \sin 2 \beta $. Mixing induced CP asymmetries
measured in many $b \to s \bar q q $ modes give the trend
$S_{s \bar q q} < \sin 2 \beta$.

In this paper we have studied the $B \to K \pi $  decay modes,
which receive dominant contributions from $b \to s$ QCD penguins in the
SM. In the SM its mixing induced CP violation parameter 
in $B^0 \to K^0 \pi^0$ is expected
to be larger than that of $\sin 2 \beta$, obtained from $b \to c \bar
c s$ transitions and $\Delta A_{CP}(K \pi) \approx
2.5 \%$. However the observed value $S_{\pi^0 K_s}$ is less than $\sin 2
\beta $ by nearly 1-sigma and $\Delta A_{CP}(K \pi)$ deviates
from its SM value by nearly 4 $\sigma$.
 Also the SM predicted branching ratio in $B^0 \to \pi^0 K^0$ is
found to be less than that of observed value. Since  the final
$\pi^0$ can materialize from a $Z'$ boson, this decay mode is quite
sensitive to electroweak penguin contributions. We have considered
the effect of an extra non-universal $Z'$ gauge boson which is
expected to give significant contributions to electroweak penguin
sector. We have shown that the observed anomalies in this mode could
be successfully explained with a TeV range $Z'$ boson.
In future with improved statistics, this mode can provide an
indirect hint for the existence of $Z'$ boson.

\acknowledgments The work of RM was partly supported by Department
of Science and Technology, Government of India, through grant Nos.
SR/S2/HEP-04/2005 and SR/S2/RFPS-03/2006. AG would like to thank
Council of Scientific and Industrial Research and Department of
Science and Technology, Government of India, for financial support.

\end{document}